\documentclass[twocolumn,prl,aps,showpacs,preprintnumbers,superscriptaddress]{revtex4}

\usepackage{graphicx}

\newcommand{\be}{\begin{equation}}
\newcommand{\ee}{\end{equation}}
\newcommand{\bea}{\begin{eqnarray}}
\newcommand{\eea}{\end{eqnarray}}
\newcommand{\HH}{{\cal H}}

\newcommand{\la}{\langle}
\newcommand{\ra}{\rangle}

\newcommand{\lp}{\left(}
\newcommand{\rp}{\right)}

\begin{document}
\title{Coherent Quasiclassical Dynamics of a Persistent Current Qubit}

\author{D.~M.~Berns}
 \email{dmb@mit.edu}
 \affiliation{Department of Physics, Massachusetts Institute of Technology, Cambridge MA 02139}
\author{W.~D.~Oliver}
 \affiliation{MIT Lincoln Laboratory, 244 Wood Street, Lexington, MA 02420}
\author{S.~O.~Valenzuela}
 \affiliation{MIT Francis Bitter Magnet Laboratory, Cambridge, MA 02139}
\author{A.~V.~Shytov}
 \affiliation{Physics Department, Brookhaven National Laboratory, Upton, NY 11973-5000}
\author{K.~K.~Berggren}
 \altaffiliation[Present address: ]{EECS Department, MIT} 
 \affiliation{MIT Lincoln Laboratory, 244 Wood Street, Lexington, MA 02420}
\author{L.~S.~Levitov}
 \affiliation{Department of Physics, Massachusetts Institute of Technology, Cambridge MA 02139}
\author{T.~P.~Orlando}
 \affiliation{Department of Electrical Engineering and Computer Science, Massachusetts Institute of Technology, Cambridge, MA 02139}


\begin{abstract}
A new regime of coherent quantum dynamics of a qubit is realized
at low driving frequencies in the strong driving limit. Coherent
transitions between qubit states occur via the Landau-Zener
process when the system is swept through an energy-level avoided
crossing. The quantum interference mediated by repeated
transitions gives rise to an oscillatory dependence of the qubit
population on the driving field amplitude and flux detuning. These
interference fringes, which at high frequencies consist of
individual multiphoton resonances, persist even for driving
frequencies smaller than the decoherence rate, where individual
resonances are no longer distinguishable. A theoretical model that
incorporates dephasing agrees well with the observations.
\end{abstract}
\maketitle
 \vspace{-10mm}

Macroscopic quantum systems coherently driven by external fields
provide new insights into the fundamentals of quantum mechanics
and hold promise for applications such as quantum
computing~\cite{Mooij05}. Superconducting Josephson devices are
model quantum systems
that can be manipulated by RF driving fields~\cite{Makhlin01a},
and recent years have seen rapid progress in the understanding of
their quantum
dynamics~\cite{Nakamura99a,Friedman00a,Wal00a,Nakamura01,Vion02a,Yu02a,Martinis02a,Chiorescu03a}.
Quantum coherence of these systems can be probed by temporal Rabi
oscillations~\cite{Nakamura99a,Nakamura01,Vion02a,Yu02a,Martinis02a,Chiorescu03a}.
There, the driving-field frequency $\nu$ equals the energy level
separation $\Delta E$, and the population of the two levels
oscillates at a frequency $\omega_{\text{R}}$ much smaller than
$\Delta E$. In the weak driving limit, $\hbar \omega_{\text{R}}
\approx A \ll \Delta E = h\nu$, where $A$ is the driving amplitude
parameterized in units of energy.

Coherent quantum dynamics can also be investigated
at driving frequencies much less than $\Delta E$, and at strong
driving amplitude $A\approx \Delta E\gg h\nu$.
In this case, the transitions occur via the Landau-Zener (LZ)
process at a level crossing~\cite{Shytov03,Izmalkov04}. Acting as
a coherent beamsplitter, LZ transitions create a quantum
superposition of the ground and excited states and, upon
repetition, induce quantum mechanical interference. The latter
leads to Stueckelberg-type
oscillations~\cite{Stueckelberg32a,H_Nakamura01} in analogy to a
Mach-Zehnder (MZ) interferometer~\cite{Oliver05,Sillanpaa05}.
These oscillations are also related to photoassisted
transport~\cite{Tien63a,Kouwenhoven94,NakamuraTsai99} and Rabi
oscillations observed in the multiphoton
regime~\cite{Nakamura01,Saito06a}. MZ-type interference is a
unique signature of temporal coherence complementary to Rabi
oscillations,
with
the time between sequential LZ transitions clocking the dynamics
similarly to Rabi pulse width.

\begin{figure}[t]
\includegraphics[width=3.4in]{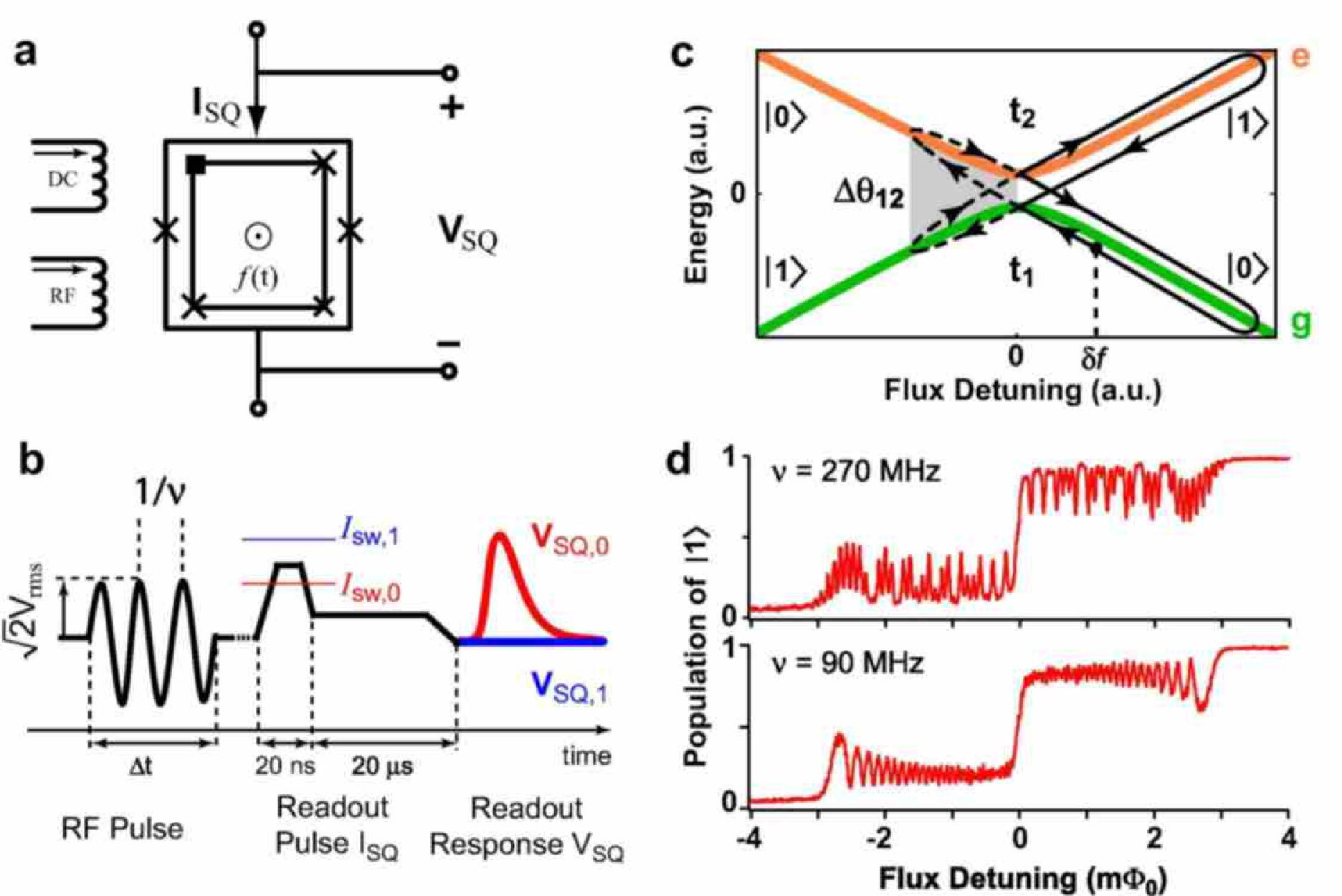}
\caption[t]{(a) Schematic of the PC qubit surrounded by a DC SQUID
readout. DC and RF fields control the state of the qubit. (b) An
RF pulse of duration $\Delta t\gg \nu^{-1}$ drives the qubit, and
its state is inferred from the voltage $V_{\rm SQ}$ across the
SQUID pulsed with current $I_{\rm SQ}$. (c) The qubit experiences
two Landau-Zener transitions over a single RF period, accumulating
a relative phase $\Delta\theta_{12}$ between them. (d) The
resulting interference fringes in qubit population for $\nu=270
\,\,{\rm and}\,\, 90\,{\rm MHz}$, and $V_{\text{rms}}=240 \,\,{\rm
and}\,\, 171\,{\rm mV}$ respectively (vertical lines on
Fig.\ref{fig2}). }
 \label{fig1}
\vspace{-5mm}
\end{figure}

In this Letter, we report a new {\it quasiclassical} regime
which exhibits coherence even at driving frequencies low compared
to dephasing rate, $\nu T_2 \lesssim 1$~\cite{T2nu<1}.
This occurs because the interval between consecutive LZ
transitions, relevant for MZ interference, is only a fraction of
the driving field period.  We investigate the crossover between
the multiphoton and quasiclassical regimes,
demonstrating that coherent MZ-type interference fringes in the
qubit population persist for frequencies $\nu T_2 \lesssim 1$
even though individual multiphoton resonances can no longer be
resolved. This behavior should be contrasted with Rabi
oscillations, where at low driving frequency, $\nu T_2\lesssim 1$,
there is no signature of coherence. The crossover
between the two regimes, $\nu T_2\sim 1$, is also influenced by 
inhomogeneous broadening, as discussed below.


In our experiment we utilize a persistent-current (PC)
qubit~\cite{Orlando99a}: a superconducting loop interrupted by
three Josephson junctions (JJ), one of which has a reduced
cross-sectional area (Fig.\ref{fig1}a). A time-dependent magnetic
flux $f(t) = f^{\text{dc}} + f^{\text{ac}}$ controls the qubit.
For $f(t) \approx \Phi_0/2$, the qubit exhibits a double-well
potential profile with individual wells representing diabatic
circulating-current states, $|0\ra$ and $|1\ra$, with energies
$\pm\epsilon \propto \pm \delta f$, where $\delta f \equiv
f^{\text{dc}} - \Phi_0/2$ is the flux detuning. These states are
coupled with a tunneling energy $\Delta$. The driving and readout
pulse sequence is illustrated in Fig.\ref{fig1}(b). Qubit
transitions are driven by a microwave flux $f^{\text{ac}} \propto
A \cos 2\pi\nu t$, with $A$, parameterized in units of energy,
proportional to the microwave source voltage $V_{\text{rms}}$. The
qubit state is read out with a DC SQUID, whose switching current
$I_{\text{SW}}$ depends on the flux generated by the qubit and,
thereby, the qubit circulating-current state. The device was
fabricated at Lincoln Laboratory using a fully-planarized niobium
trilayer process and optical lithography. The device has a
critical current density $J_{\rm c} \approx 160 \,{\rm A/cm^2}$,
and the characteristic Josephson and charging energies are $E_{\rm
J} \approx (2\pi\hbar)300\,{\rm GHz}$ \,\,{\rm and}\,\, $E_{\rm C}
\approx (2\pi\hbar)0.65\,{\rm GHz}$ respectively. The ratio of the
qubit JJ areas is $\alpha \approx 0.84$, and $\Delta\approx
(2\pi\hbar)10\,{\rm MHz}$. The experiments were performed in a
dilution refrigerator at a base temperature of 20 mK. The device
was magnetically shielded, and all electrical lines were carefully
filtered and attenuated to reduce noise (see Ref.~\cite{Oliver05}
for details).

The qubit dynamics in the strongly driven limit is influenced by
quantum interference at sequential LZ transitions. As illustrated
in Fig.\ref{fig1}(c), the qubit is initially prepared in the
ground state at flux detuning $\delta f$, and, after a first LZ
transition at time $t_1$, it is in a coherent superposition of the
two diabatic states. For times $t_1 < t < t_2$, the superposition
state accumulates a relative phase $\Delta\theta_{12}$, which
mediates the quantum interference at the second LZ transition at
time $t_2$. The sequence of two LZ transitions, repeated many
times during the RF pulse, is analogous to a cascade of MZ
interferometers. One expects MZ-type interference fringes in the
qubit population due to changes in $\Delta\theta_{12}$ associated
with changes in $V_{\text{rms}}$ and $\delta f$, which are indeed
observed (Fig.\ref{fig1}d).

\begin{figure}[t]
\includegraphics[width=3.2in]{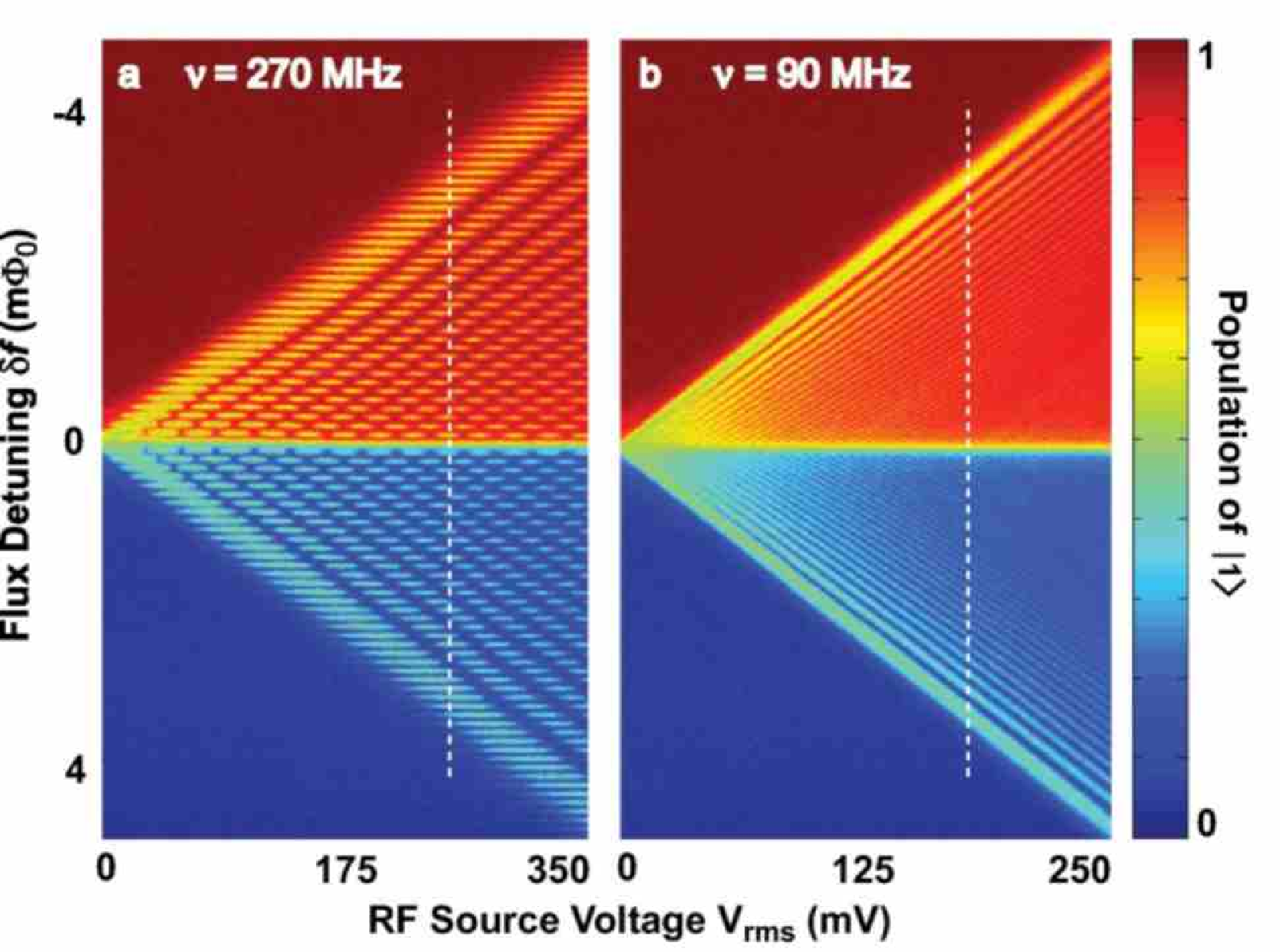}
\caption[t]{
Measured qubit population at strong driving in two regimes.
(a) $\nu = 270\,{\rm MHz}$.  Multiphoton resonances of order up
to $n=45$ can be discerned ($\nu T_2> 1$). (b) $\nu = 90\,{\rm MHz}$.
Individual resonances are no longer distinguishable ($\nu
T_2\lesssim 1$), but coherent interference is still observed.
Vertical lines indicate the scans displayed in Fig.\ref{fig1}d.
A pulse of duration $\Delta t=3\,{\rm \mu s}$ was used in both
cases.
}
 \label{fig2}
\vspace{-5mm}
\end{figure}

Figure \ref{fig2} presents the measured qubit population of state
$|1\rangle$ (color scale) as a function of $V_{\text{rms}}$ and
$\delta f$ for high- and low-frequency driving, $\nu = 270
\,\,{\rm and}\,\, 90\,{\rm MHz}$ respectively. Population transfer
due to qubit driving appears at $V_{\text{rms}}$ exceeding a
threshold value which varies linearly with $|\delta f|$ and
symmetrically about the qubit step.
For high-frequency driving, $\nu T_2\gtrsim 1$, the individual
multiphoton resonances are distinguishable and form a ``Bessel
ladder''~\cite{Oliver05}(Fig.~\ref{fig2}a). The population of
state $|1\rangle$ for the $n^{\text{th}}$-photon resonance follows
a Bessel-function dependence, $J_n^2(A/h\nu)$. The range of
$\delta f$ in Fig. 2a accommodates photon transitions with
$n=1-45$, which together define coherent MZ interference-fringe
bands of discrete resonances.

In contrast, for low-frequency driving, $\nu T_2\lesssim 1$, the
individual photon resonances are no longer distinguishable because
the resonance widths exceed the resonance spacing
(Fig.~\ref{fig2}b). Nonetheless, the MZ interference-fringe bands,
a signature of coherence in the strongly driven regime, indicate
that the coherent interference mediating the population transfer
persists.

To understand these results we consider a driven qubit subject to
the effects of decoherence ($\hbar=1$):
\be \HH = -\frac12 \lp \matrix{h(t) & \Delta \cr \Delta &
-h(t)}\rp . \label{Eq:Hamiltonia} \ee
Here, $h(t)=\epsilon+\delta\epsilon(t)+A\cos 2\pi\nu t$ is the
energy detuning from an avoided crossing modulated by the driving
field in the presence of classical noise $\delta\epsilon(t)$. By a
gauge transformation, the Hamiltonian is brought to the form
\be
\HH =
-\frac12 \lp \matrix{ 0 & \Delta(t) \cr \Delta^\ast(t) & 0}\rp
,\quad
\Delta(t)=\Delta e^{-i\phi(t)}
,
\ee
where $ \phi(t)=\int^t_0 h(\tau)d\tau $. Perturbation theory gives
the rate of LZ transitions between the states $|0\ra$ and $|1\ra$:
\be\label{eq:pert_theory}
W = \lim_{\delta t\to\infty} |A_{t,t'}|^2/\delta t
,\quad
A_{t,t'}=\int_t^{t'}\Delta(t')dt'
,
\ee
where $\delta t=t'-t>0$, and the limit physically means that
$\delta t$ is large compared to $T_2$.

For the perturbation approach to be valid, the change of qubit
population must be slow on the scale of $T_2$. This condition can
be written down as $W\ll \Gamma_2=1/T_2$. We stress that this
inequality does not imply that the effect of driving the qubit is
weak. The rate $W$ can still be large compared to the inelastic
relaxation rate $\Gamma_1$, leading to the strong deviation of
population from equilibrium observed in our experiment.

To evaluate $W$, we write the expression in
Eq.(\ref{eq:pert_theory}) as
\be\label{eq:Wgeneral}
    W =
    \frac14 \int \la \Delta(t+\tau) \Delta^\ast(t)\ra d\tau
    , \quad
    \Delta(t)=\Delta e^{-i\phi(t)}
    .
\ee
By introducing Bessel functions in the Fourier series of $e^{-i(A/\omega)\sin\omega t}$, $\omega=2\pi\nu$,
we have
\[ \nonumber
    e^{-i\phi(t)}=e^{-i\epsilon t-i\delta\phi(t)}\sum_n J_n(x)e^{i\omega n t}
    ,\quad
    x=\frac{A}{\omega}=\frac{A}{h\nu}
    .
\]
We average over $\delta\phi(t)$ with the help of the white noise
model $\la
e^{i\delta\phi(t)-i\delta\phi(t')}\ra=e^{-\Gamma_2|t-t'|}$, and
integrate in (\ref{eq:Wgeneral}) as $\int e^{-i(\epsilon-\omega
n)\tau -\Gamma_2|\tau|}d\tau=2\Gamma_2/((\epsilon-\omega
n)^2+\Gamma_2^2)$ to obtain
\be\label{eq:W}
    W(\epsilon,A)
    = \frac{\Delta^2}2\sum_n\frac{\Gamma_2
    J_n^2(x)}{(\epsilon-\omega n)^2+\Gamma_2^2}
    .
\ee
For large $n$, Bessel functions can be expressed through the Airy
function $\text{Ai}(u)=\frac1{\pi}\int_0^\infty \cos(uy+\frac13
y^3)dy$ as $J_n(x)=a\text{Ai}(a(n-x))$, $a=(2/x)^{1/3}$. Using the
identity $\cot z =\sum (z-\pi n)^{-1}$ we approximate
Eq.\,(\ref{eq:W}) as
\be\label{eq:Wairy}
    W = \frac{\pi a^2\Delta^2}{2\omega} \text{Im}\,\cot\lp
    \frac{\pi}{\omega}(\epsilon-i\Gamma_2)\rp \text{Ai}^2\lp
    \frac{a}{\omega}(\epsilon-A)\rp
    .
\ee
There are two main regimes exhibited by this expression:
 (i) $\nu \gtrsim \Gamma_2$, and
 (ii) $\nu \lesssim \Gamma_2$.
 In case (i),
we have a sum of non-overlapping resonances. For each value of
$\epsilon$, the sum is dominated by the term with $n$ the nearest
integer to $\epsilon/\omega$, giving rise to resonances of
strength $J_n^2(x)$, the
 Bessel ladder of Ref.~\cite{Oliver05}.

In contrast, in case (ii), the peaks in Eq.~(\ref{eq:W}) are
overlapping. Setting $\cot=i$ in
Eq.(\ref{eq:Wairy})~\cite{T2nu<1}, we obtain
\be\label{eq:low omega}
    W(\epsilon,A) \approx \frac{\pi
    a^2\Delta^2}{2\omega}
    \text{Ai}^2(a(\epsilon-A)/\omega)
    .
\ee
The effect of $\Gamma_2$ on the Airy function oscillation is small
at $\Gamma_2\lesssim (2\pi/a)\nu$. Since $a\approx 0.3$ for
$\epsilon/h\nu\lesssim 50$, this condition is compatible with
$\nu\lesssim \Gamma_2$. Eq.(\ref{eq:low omega}) can also be
obtained by considering just two subsequent passages of a level
crossing at a short time separation $|t_2-t_1|\ll \nu^{-1}$, and
ignoring the periodicity of the driving.

Since $\text{Ai}(u<0)$ oscillates as
$\pi^{-1/2}|u|^{-1/4}\cos(\frac23|u|^{3/2}-\frac{\pi}4)$, while
$\text{Ai}(u>0)$ decays exponentially, Eq.(\ref{eq:low omega})
implies that the transitions occur only for $A\gtrsim \epsilon$,
with a rate which oscillates as a function of $A-\epsilon$. The
oscillations are the same for both integer and noninteger
$\epsilon/h\nu$, confirming that, while the resonances merge into
a continuous band, the interference fringes persist at $\nu
\lesssim \Gamma_2$, in agreement with our observations.


\begin{figure}[t]
\includegraphics[width=2.8in]{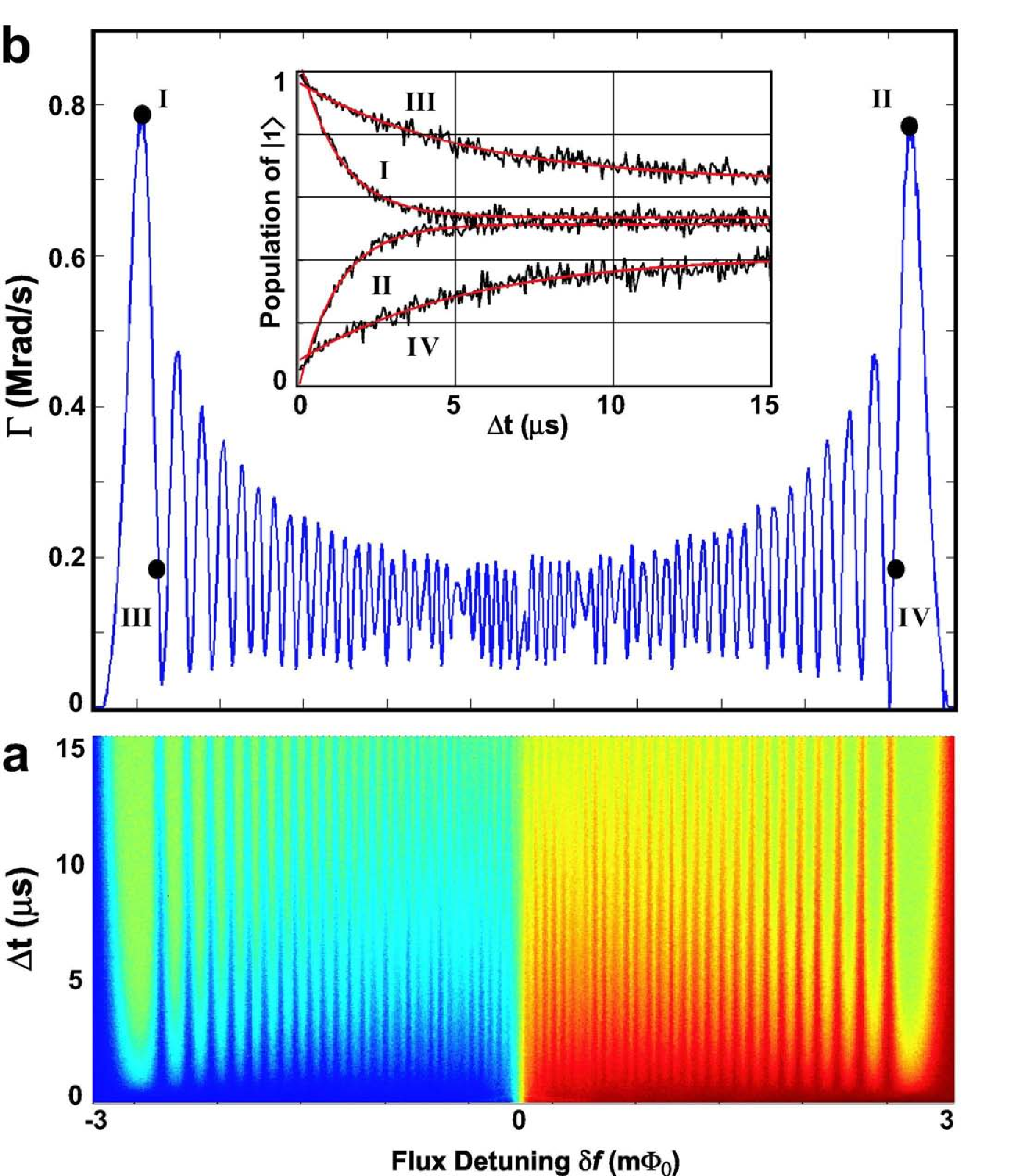}
    \caption[t]{(a) Time evolution of excited state population ($\nu =
    90\,\text{MHz}$, $V_{rms} = 171\,\text{mV}$)
    obtained by varying the pulse width $\Delta t$.
    (b) The
    characteristic rate $\Gamma$ as a function of flux detuning $\delta
    f$ obtained by fitting to the exponential time dependence [Eq.(\ref{eq:m
    time})]
    (inset shows examples of fits for the points I, II, III and IV).
    }
    \label{fig3}
\vspace{-5mm}
\end{figure}

To describe the population dynamics in the presence of driving, we
employ a rate equation approach, in which the qubit level occupations
obey $\dot p_i=\sum_j g_{ij}p_j$, where
\be
g_{01}=-g_{11}=W+\Gamma_1
,\quad
g_{10}=-g_{00}=W+\Gamma'_1
.
\ee
Here, $\Gamma_1=1/T_1$, $\Gamma'_1=\Gamma_1e^{-\beta\epsilon}$ are
the down and up relaxation rates. The magnetization of the
stationary state is
$m_s=p_0-p_1=(\Gamma_1-\Gamma'_1)/(2W+\Gamma_1+\Gamma'_1)$, which
gives the equilibrium value $m_0=\tanh \frac12 \beta\epsilon$ at
weak excitation, and $m_s\ll m_0$ at high excitation.

To validate this model, we investigate the interference fringes in
the excited state population as a function of RF pulse length
$\Delta t$ (Fig.\ref{fig3}a). The rate equation predicts an
exponential time dependence for the magnetization,
%
\be\label{eq:m time}
    m(\Delta t)=m_s+(m_0-m_s) e^{- \Gamma \Delta t} ,\quad
    \Gamma = 2W+\Gamma_1+\Gamma'_1
    .
\ee
By fitting exponentials [Eq.(\ref{eq:m time})] to the qubit
population at each flux detuning, we find the rate $\Gamma$ which
characterizes how fast the stationary state is approached
(Fig.\ref{fig3}b). Since our $T_1 \approx 20\,\mu {\rm
s}$~\cite{Oliver05} is much longer than the observed transition
time, we have $\Gamma \approx 2W$.
Comparing the extracted $\Gamma$ with Eq.(\ref{eq:low omega}), we
obtain $\Delta /2\pi\hbar = 13\,{\rm MHz}$.

\begin{figure}[t]
\includegraphics[width=3in]{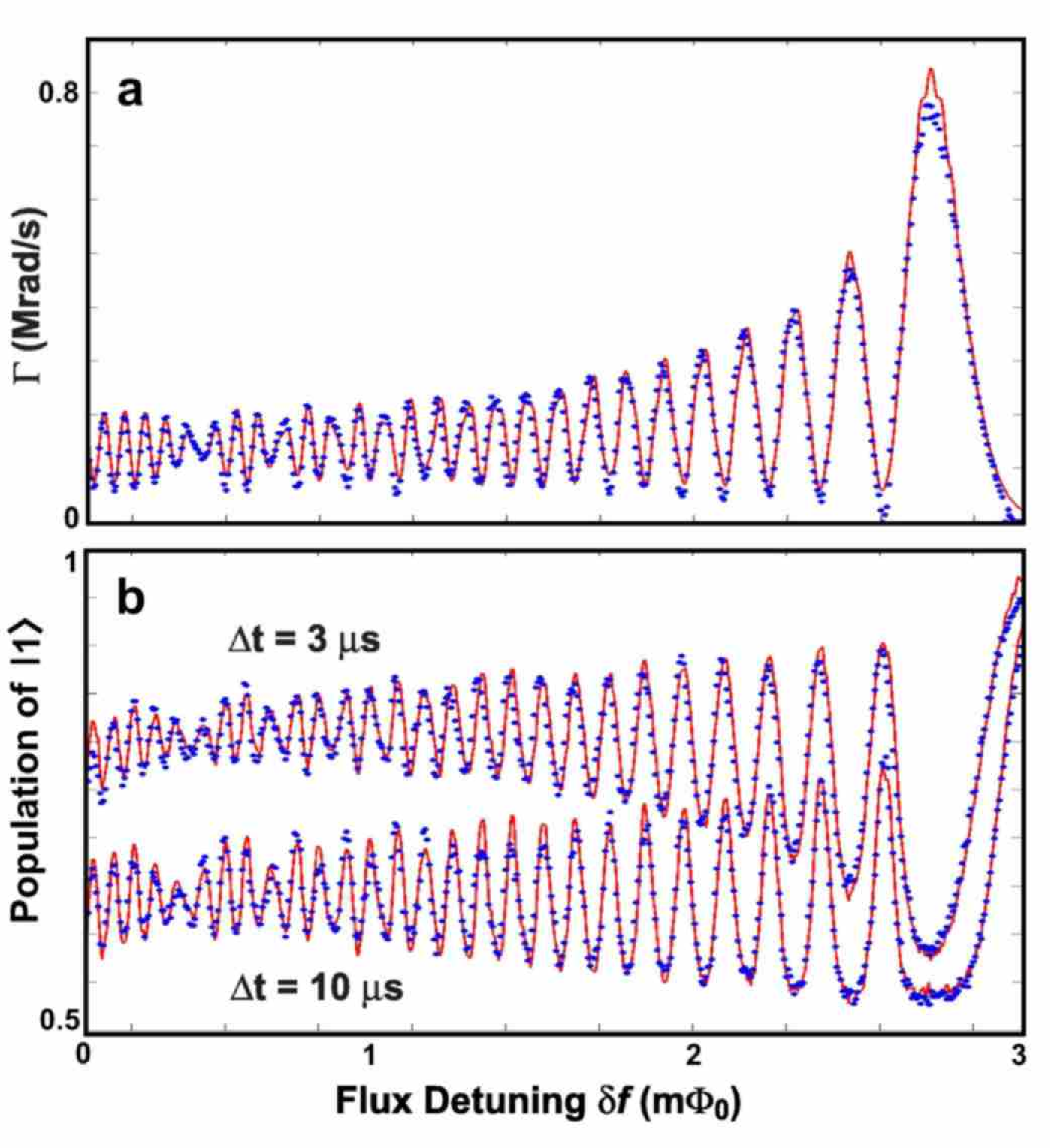}
    \caption[t]{Comparison of experiment (blue) and theory (red).
    (a) The transition rate from the right half of Fig.\ref{fig3}b
    fitted with $\Gamma$ defined by Eqs.(\ref{eq:W}),(\ref{eq:m time}).
    (b) State $|1\rangle$ occupation taken from Fig.\ref{fig3}a,
    compared to the model, Eq.(\ref{eq:m time}). }
 \label{fig4}
 \vspace{-4mm}
\end{figure}

In Fig.~\ref{fig4} we compare the experimental characteristic rate
and qubit population with those predicted by the model
[Eqs.(\ref{eq:W}),(\ref{eq:m time})]. From the best fits
 we obtain
 $\Gamma_2 /2 \pi  = 12-18\,{\rm MHz}$
($T_2\approx 9-13\,{\rm ns}$), consistent with the transition
between the multiphoton and quasiclassical regimes of
Fig.~\ref{fig2}: $270 \, \text{MHz} > \Gamma_2=1/T_2 \gtrsim 90 \,
\text{MHz}$.

\begin{figure}[h]
\includegraphics[width=3.2in]{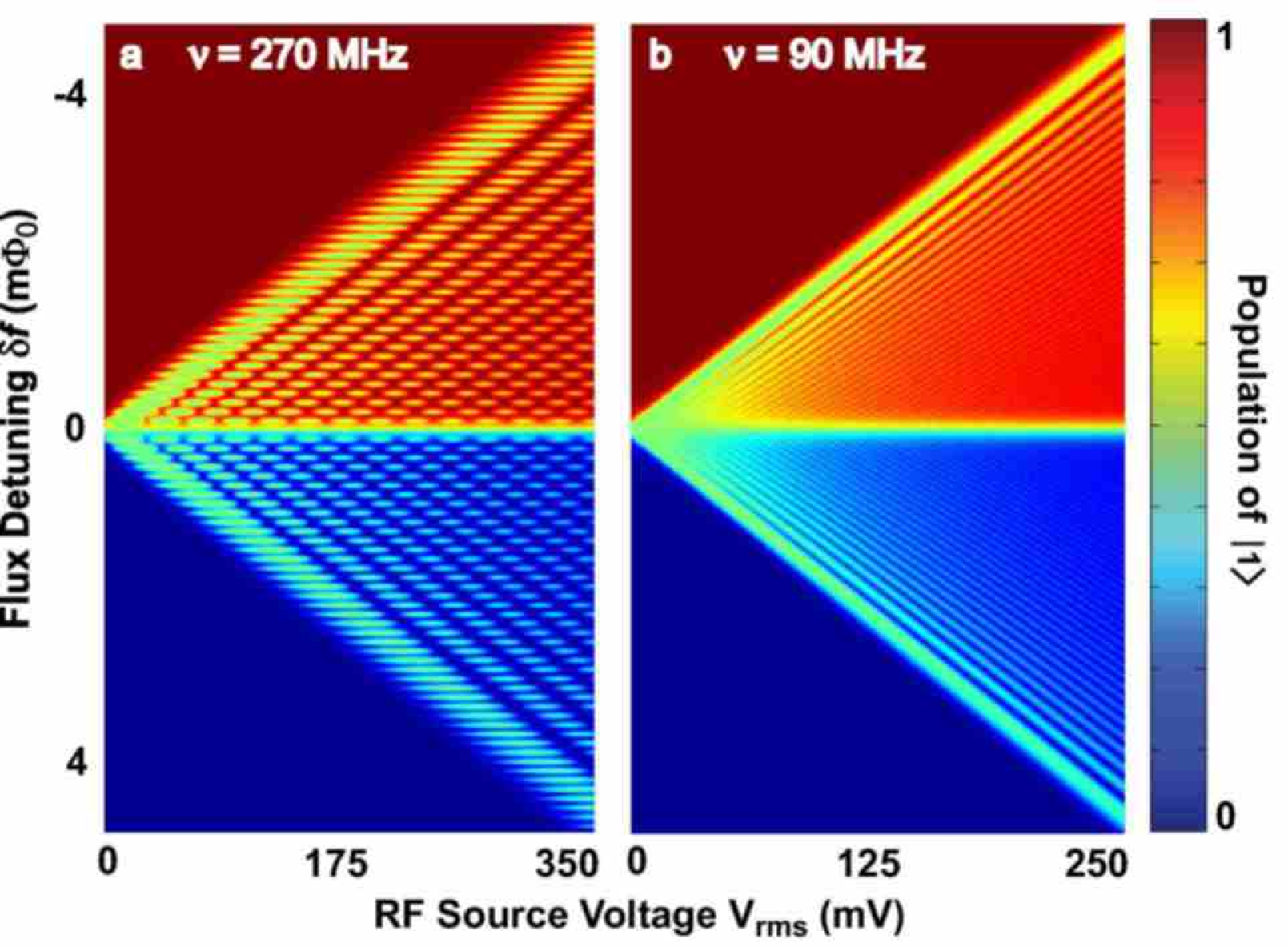}
    \caption[t]{Simulation of qubit population using model parameters extracted from
    data. (a) $\nu = 270\,{\rm MHz}$. (b) $\nu = 90\,{\rm MHz}$.}
    \label{fig5}
\vspace{-2mm}
\end{figure}

Inhomogeneous broadening is incorporated into the model by
assuming a Gaussian broadening mechanism with standard deviation
$\sigma / 2 \pi = 40-45\,{\rm MHz}$. The resulting power-broadened
linewidth is approximately $150 \text{ MHz}$, consistent with the
linewidth observed in Fig.~\ref{fig2}a. Best fits in
Fig.~\ref{fig4} are obtained with slightly different values of
$\Gamma_2$ and $\sigma$ within the ranges above. By using the fit
parameters for the 3$\mu s$ magnetization curve, we can calculate
the qubit population in the multiphoton (Fig.~\ref{fig5}a) and
quasiclassical (Fig.~\ref{fig5}b) regimes as a function of $\delta
f$ and $V_{\rm rms}$.

In conclusion, we have observed quantum coherent qubit dynamics at
strong driving for frequencies smaller than the dephasing rate. In
this limit, well-resolved multiphoton transitions merge into a
continuous band, while the Mach-Zehnder-like coherent interference
pattern persists. A simple model of a driven two-level system
subject to decoherence is in remarkable agreement with the
observed interference patterns.

We thank V. Bolkhovsky, G. Fitch, D. Landers, E.
Macedo, R. Slattery, and T. Weir at MIT Lincoln Laboratory for
fabrication and technical assistance; D. Cory, A.~J. Kerman, and
S. Lloyd for helpful discussions. This work was supported by AFOSR
(F49620-01-1-0457) under the DURINT program. The work at Lincoln
Laboratory was sponsored by the US DoD under Air Force Contract
No. FA8721-05-C-0002. AVS acknowledges support by US DOE under
contract No. DEAC 02-98 CH 10886.


\begin{thebibliography}{10}
\bibitem{T2nu<1} We emphasize that the crossover between high
and low frequencies in this case occurs at $\nu T_2\sim 1$ rather than at
$\omega T_2\sim 1$. This can be inferred from the expression
$\text{Im}\,\cot(\pi(\epsilon-i\Gamma_2)/\omega)$, Eq.(\ref{eq:Wairy}),
which describes multiphoton resonances broadening. At large $\Gamma_2$
this gives $1+2e^{-2\pi\Gamma_2/\omega}\cos(2\pi\epsilon/\omega)
+O(e^{-4\pi\Gamma_2/\omega})$.
Multiphoton oscillations become dephased
at $\nu\approx\Gamma_2$.
\end{thebibliography}
\end{document}